\documentclass[prl,twocolumn,showpacs,preprintnumbers,amsmath,amssymb]{revtex4}

\usepackage{graphicx}
\usepackage{dcolumn}
\usepackage{bm}
\usepackage{amsmath}

\begin{document}


\title{Generation of two-mode field squeezing through selective dynamics in cavity QED}

\author{Pengbo Li}

\affiliation{State Key Laboratory for Mesoscopic Physics, Department
of Physics, Peking University, Beijing 100871, China }

\date{\today}

\begin{abstract}
We propose a scheme for the generation of a two-mode field squeezed
state in cavity QED. It is based on two-channel Raman excitations of
a beam of three-level atoms with random arrival times by two
classical fields and two high-Q resonator modes. It is shown that by
suitably choosing the intensities and detunings of fields the
dynamical processes can be selective and two-mode squeezing between
the cavity modes can be generated at steady state. This proposal
does not need the preparation of the initial states of atoms and
cavity modes, and is robust against atomic spontaneous decay.
\end{abstract}

 \maketitle

Squeezing is one of the most striking features of quantum optics,
which can be simply defined as the reduction of quantum fluctuations
in a certain quadrature below the vacuum level, at the expense of
increasing them in its canonically conjugate
variable\cite{quantum_optics}. Various theoretical schemes and
experimental protocols have been proposed or even implemented to
produce the squeezed states of electromagnetic
field\cite{Prl-59,Prl-88-093601,Prl-91-103601}. Recently with the
advent of quantum information and
communication\cite{quantum_information}, squeezed states of light
have played very important roles in numerous quantum information
protocols\cite{RMP-77-513}, e.g., the realizations of continuous
variable computation \cite{Prl-82-1784} and
teleportation\cite{SCi282}. Also two-mode squeezed states
\cite{pra-31-1985} can lead to efficient distribution of
entanglement and implementation of quantum channels by improving the
low squeeze parameters\cite{Prl-90-047905}. Two-mode squeezing has
already been realized through Kerr nonlinearity in optical fibers
and with atomic clouds in optical cavities\cite{Prl-92-123601}. In
the context of cavity QED\cite{Kimble,Sci298}, two-mode field
squeeze operators in optical cavities with atomic ensembles have
been proposed\cite{Prl-96-010502}. Also two-mode squeezing of
separated atomic ensembles has been presented\cite{Prl-96-053602}.
Most recently, a scheme for generating two-mode field squeezing has
been proposed\cite{prl-98}, which is based on atomic reservoir in
four-wave mixing processes in cavity QED. However, to implement this
protocol, one has to prepare the two-level Rydberg atoms in a
coherent superposition of ground state and excited state before the
atoms enter the cavity.

In this paper, we propose a scheme for the generation of a two-mode
field squeezed state in cavity QED. It does not need neither the
preparation of the initial state of the atoms nor the initial state
of the cavity. The whole system has only to stay in the ground
states initially. This proposal is based on two-channel Raman
excitations \cite{Prl-96-010502,Law} of a beam of three-level atoms
with random arrival times by two classical fields and two high-Q
cavity modes. This process corresponds to a form of atomic reservoir
engineering, where the resonator is pumped randomly by a beam of
atoms which constitute a spin reservoir\cite{Englert}. We show that
by suitably choosing the intensities and detunings of fields the
dynamical processes can be selective, which is utilized to generate
two-mode squeezing between the cavity modes at steady state. To
implement this scheme it does not require atomic detection nor
velocity selection, and is robust against atomic spontaneous decay.
With presently available experimental setups in cavity QED this
protocol can be realized.

Our proposal relies on the two-channel Raman excitations of a beam
of three-level $\Lambda$ configuration
atoms\cite{Prl-96-010502,Law}.
\begin{figure}[h]
\centering
\includegraphics[bb=105 527 430 751,totalheight=1.5in,clip]{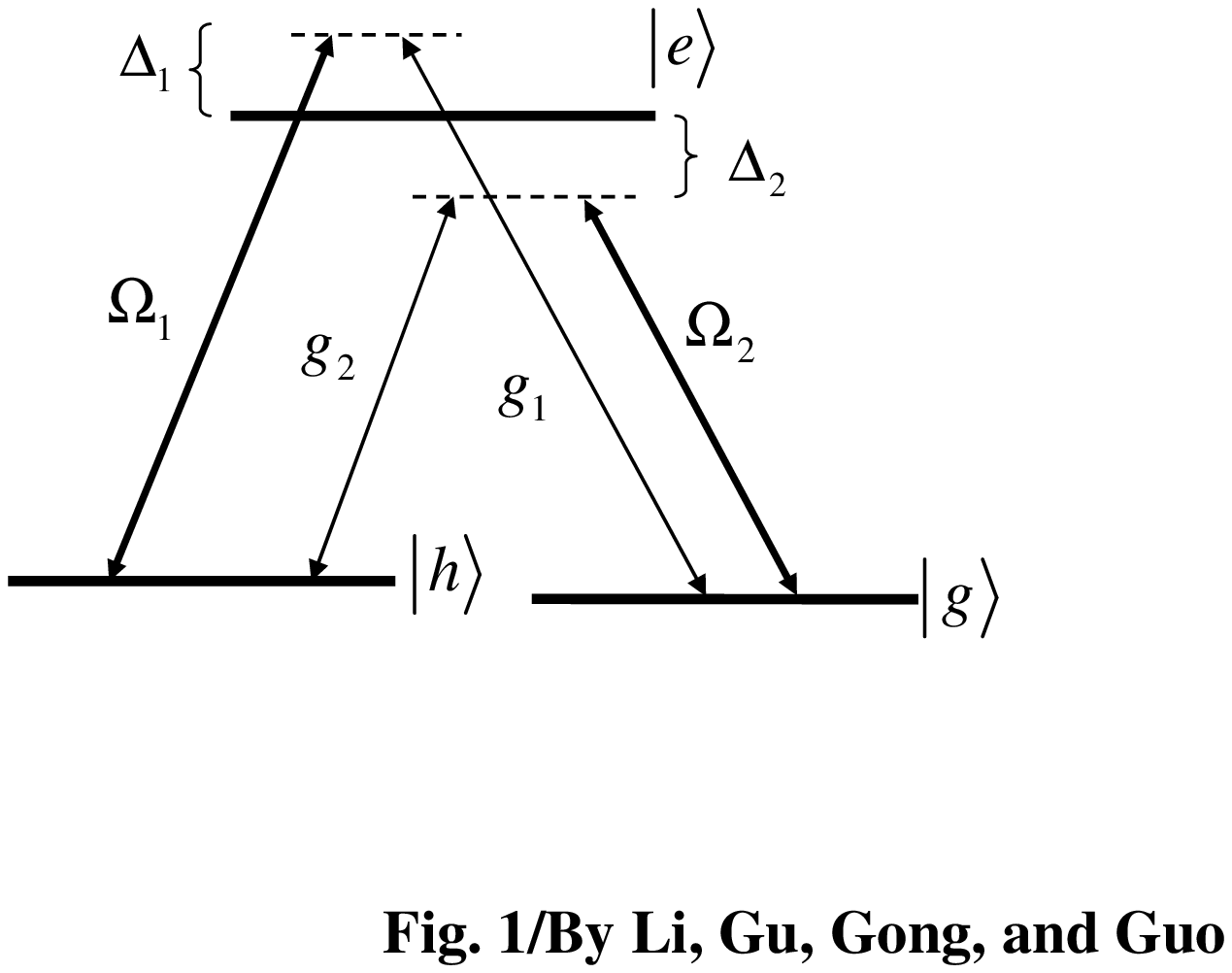}
\caption{ Scheme for the two-channel Raman excitation of the
three-level atoms.}
\end{figure}
As sketched in Fig. 1, two classical fields of frequencies
$\omega_1$ and $\omega_2$ drive dispersively each three-level atom,
establishing a couple of Raman laser system through two high-Q
cavity modes of frequencies of $\nu_1$ and $\nu_2$. The ground state
of the atoms is labeled as $\vert g\rangle$, the metastable state as
$\vert h\rangle$, and the excited state as $\vert e\rangle$. The
classical fields drive dispersively the transitions $\vert
h\rangle\leftrightarrow \vert e\rangle$ and $\vert
g\rangle\leftrightarrow \vert e\rangle$ with Rabi frequencies
$\Omega_1$ and $\Omega_2$. The cavity modes couple the transitions
$\vert g\rangle\leftrightarrow \vert e\rangle$ and $\vert
h\rangle\leftrightarrow \vert e\rangle$ with coupling constants
$g_1$ and $g_2$. The detunings for these transitions are
$\Delta_1=\omega_{eh}-\omega_1=\omega_{eg}-\nu_1$, and
$\Delta_2=\omega_{eg}-\omega_2=\omega_{eh}-\nu_2$. In the
interaction picture, the associated  Hamiltonian under the dipole
and rotating wave approximation is given by (let $\hbar=1$)
\begin{equation}
H_I=\Omega_1\hat{\sigma}_{eh}e^{i\Delta_1t}+\Omega_2\hat{\sigma}_{eg}e^{i\Delta_2t}+g_1\hat{a}_1\hat{\sigma}_{eg}e^{i\Delta_1t}+
g_2\hat{a}_2\hat{\sigma}_{eh}e^{i\Delta_2t}+H.c.,
\end{equation}
where $\hat{\sigma}_{jm}=\vert j\rangle\langle m\vert$ is the atomic
transition operator, and $\hat{a}_i $ is the annihilation operator
for the mode with frequency $\nu_i (i=1,2)$. We consider dispersive
detunings
$|\Delta_1|,|\Delta_2|,|\Delta_1-\Delta_2|\gg|\Omega_1|,|g_1|,|\Omega_2|,|g_2|$.
Since level $\vert e\rangle$ is coupled dispersively with both
levels $\vert g\rangle$ and $\vert h\rangle$, it can be
adiabatically eliminated and atomic spontaneous emission can be
neglected\cite{James}. Then we obtain the effective Hamiltonian
describing the two-channel Raman excitations of the atoms
\begin{eqnarray}
H_{eff}=&&(\frac{|\Omega_1|^2}{\Delta_1}+\frac{|g_2|^2}{\Delta_2}\hat{a}^{\dag}_2\hat{a}_2)\hat{\sigma}_{hh}+
(\frac{|\Omega_2|^2}{\Delta_2}+\frac{|g_1|^2}{\Delta_1}\hat{a}^{\dag}_1\hat{a}_1)\hat{\sigma}_{gg}\nonumber\\
&&+(\frac{\Omega_1g_1^*}{\Delta_1}\hat{a}_1^\dag+\frac{\Omega^*_2g_2}{\Delta_2}\hat{a}_2)\hat{\sigma}_{gh}
+(\frac{\Omega^*_1g_1}{\Delta_1}\hat{a}_1+\frac{\Omega_2g^*_2}{\Delta_2}\hat{a}^\dag_2)\hat{\sigma}_{hg}
\end{eqnarray}
The first two terms correspond to dynamical energy shifts of levels
$\vert g\rangle$ and $\vert h\rangle$, and the last two terms
describe transitions between these levels, accompanied by creation
or annihilation of a photon in the respective cavity mode. In the
following we assume that $\Omega_1$, $\Omega_2$, $g_1$, and $g_2$
are real for simplicity.

This effective Hamiltonian can be rewritten as
\begin{equation}
H_{eff}=H_0+(\Theta_2\hat{a}_2^\dag-\Theta_1\hat{a}_1)\hat{\sigma}_{hg}+H.c.,
\end{equation}
where
$H_0=(|\frac{g_2^2}{\Delta_2}|\hat{a}^{\dag}_2\hat{a}_2-|\frac{\Omega_1^2}{\Delta_1}|)\hat{\sigma}_{hh}+
(|\frac{\Omega_2^2}{\Delta_2}|-|\frac{g_1^2}{\Delta_1}|\hat{a}^{\dag}_1\hat{a}_1)\hat{\sigma}_{gg}$,
$\Theta_i=|\frac{\tilde{\Omega}_i}{\Delta_i}|$, and
$\tilde{\Omega}_i=\Omega_ig_i$. Using the two-mode squeezing
operator
$S_{12}(\epsilon)=exp(\epsilon^\ast\hat{a}_1\hat{a}_2-\epsilon\hat{a}_1^\dag\hat{a}_2^\dag)$
we can bring the Hamiltonian (3) to the second order
anti-Jaynes-Cummings Hamiltonian $H=H_0+H_{1}$, with
$H_0=(|\frac{g_2^2}{\Delta_2}|\hat{b}^{\dag}_2\hat{b}_2-|\frac{\Omega_1^2}{\Delta_1}|)\hat{\sigma}_{hh}+
(|\frac{\Omega_2^2}{\Delta_2}|-|\frac{g_1^2}{\Delta_1}|\hat{b}^{\dag}_1\hat{b}_1)\hat{\sigma}_{gg}$,
and
\begin{eqnarray}
H_{1}=&&-\Theta_b(\hat{b}_1\hat{\sigma}_{hg}+\hat{b}_1^\dag\hat{\sigma}_{gh}),\qquad
\mbox{if} \quad \Theta_1>\Theta_2,\\
H_{1}=&&\Theta_b(\hat{b}_2^\dag\hat{\sigma}_{hg}+\hat{b}_2\hat{\sigma}_{gh}),\qquad
\mbox{if} \quad \Theta_1<\Theta_2.
\end{eqnarray}
Here $\Theta_b=(\Theta_1+\Theta_2)\sqrt{(1-r)/(1+r)}$ with
$\epsilon=\tanh^{-1}r$, while the value of
$r=\frac{\Theta_1}{\Theta_2}$ if $\Theta_1<\Theta_2$, otherwise,
$r=\frac{\Theta_2}{\Theta_1}$ if $\Theta_1>\Theta_2$. The new
bosonic operators $\hat{b}_1$, $\hat{b}_2$ can be obtained from
$\hat{a}_1$, $\hat{a}_2$ by the two-mode squeezing transformation,
$\hat{b}_j=S^\dag_{12}(\epsilon)\hat{a}_jS_{12}(\epsilon)$. This
Hamiltonian describes an effective two-level atom coupled to the
cavity modes. The ratio of $\Theta_1$ to $\Theta_2$ determines to
which of the transformed modes the two-level transition couples.

To get more insight into the coupled system of the effective
two-level atom and cavity modes, we define $\vert n_1,n_2\rangle_a$,
and $\vert n_1,n_2\rangle_b$ as the eigenvectors of the number
operators $\hat{a}^\dag_j\hat{a}_j$ and $\hat{b}^\dag_j\hat{b}_j$,
respectively. The corresponding eigenvalues are
$n_j=0,1,2,...(j=1,2)$. The two bases are related by the
transformation $\vert n_1,n_2\rangle_b=S_{12}^\dag(\epsilon)\vert
n_1,n_2\rangle_a$. In particular, the vacuum state in the $b$ basis
is a two-mode squeezed state of the two cavity modes, $\vert
0,0\rangle_b=S_{12}^\dag(\epsilon)\vert
0,0\rangle_a=\sum_{n=1}^\infty\frac{(\tanh \epsilon)^n}{\cosh
\epsilon}\vert n,n\rangle_a$. The degree of squeezing is determined
by the $r$ and thus by the ratios
$\frac{|\tilde{\Omega}_1|}{|\tilde{\Omega}_2|}$ and
$\frac{|\Delta_1|}{|\Delta_2|}$. For $\Theta_1>\Theta_2$
($\Theta_1<\Theta_2$), the state $\vert g,0,0\rangle$ ($\vert
h,0,0\rangle$) is the ground state of the new Hamiltonian, i.e.,
$H\vert g,0,0\rangle=0 (H\vert h,0,0\rangle=0)$. A general two-mode
squeezed state can be realized by utilizing the coherent
displacement operator for the two modes, i.e., $\vert
\alpha_1,\alpha_2,\epsilon\rangle
=D_1(\alpha_1)D_2(\alpha_2)S_{12}^\dag(\epsilon)\vert 0,0\rangle_a$.
Here $D_i(\alpha_i)=exp(\alpha_i\hat{a}_i^\dag-\alpha_i^*\hat{a}_i)
(i=1,2).$

After obtaining the selective interaction of the coupled system, we
now show how to prepare the cavity modes in the two-mode squeezed
states through atomic reservoir engineering. This is achieved by an
effective dissipation process in the $b$ basis and needs two steps
to be implemented. Step 1: We set
$\Theta_1=|\frac{\tilde{\Omega}_1}{\Delta_{1}}|=|\frac{\tilde{\Omega}_1}{\Delta_{10}}|>\Theta_2=|\frac{\tilde{\Omega}_2}{\Delta_{2}}|=|\frac{\tilde{\Omega}_2}{\Delta_{20}}|$.
Then the atoms enter the cavity in the ground state $\vert g\rangle$
and undergo the dynamics of Eq.(4). In this case the average
excitations from mode $\hat{b}_1$ can be removed. Step 2:
Subsequently we set
$\Theta_1=|\frac{\tilde{\Omega}'_1}{\Delta'_{1}}|=|\frac{\tilde{\Omega}_2}{\Delta_{20}}|<\Theta_2=|\frac{\tilde{\Omega}'_2}{\Delta'_{2}}|=|\frac{\tilde{\Omega}_1}{\Delta_{10}}|$
and the dynamics of Eq.(5) can be selected. In this situation the
atoms enter in another ground state $\vert h\rangle$ and absorb in
average excitations from mode $\hat{b}_2$. In order to select this
dynamics one has to change the intensities of the pump fields (Rabi
frequencies) and the transition frequencies of the three-level atom
(detunings), i.e.,
$|\Delta'_1|=|\frac{\tilde{\Omega}'_1}{\tilde{\Omega}_2}||\Delta_{20}|$,
and
$|\Delta'_2|=|\frac{\tilde{\Omega}'_2}{\tilde{\Omega}_1}||\Delta_{10}|$.
The relation
$|\tilde{\Omega}_1|-|\tilde{\Omega}'_2|>|\tilde{\Omega}'_1|-|\tilde{\Omega}_2|$
has to be maintained and the sum of the two detunings in the two
steps keep constant, i.e.,
$|\Delta_1|+|\Delta_2|=|\Delta'_1|+|\Delta'_2|$.

This proposal utilizes the atomic reservoir engineering, where the
resonator is pumped by a beam of atoms with random arrival times. On
the other hand, the atoms should have a low pumping rate in order to
ensure that at most one atom is inside the cavity at a time. We
assume the weak coupling conditions, but only with respect to the
parameters of the effective two-level system, then the interaction
of a single atom with the cavity is a small perturbation. Let $\tau$
be the interaction time, with $\Theta_b \tau\ll 1$, and make all
atoms be initially in the ground state $\vert g\rangle$ in step 1
and in state  $\vert h\rangle$ in step 2. The differential change on
the density matrix $\hat{\rho}_c$ of the cavity in each step $j
(j=1,2)$ is given by \cite{Englert,prl-98}
\begin{equation}
\frac{\partial \hat{\rho}_c}{\partial t}\Big
|_{j}=-\frac{\gamma}{2}(\hat{b}^\dag_j\hat{b}_j
\hat{\rho}_c-2\hat{b}_j
\hat{\rho}_c\hat{b}^\dag_j+\hat{\rho}_c\hat{b}^\dag_j \hat{b}_j),
\end{equation}
where $\gamma=r_a \Theta^2_b\tau^2$ and $r_a$ is the atomic arrival
rate. Therefore, in each step $j$ we have
$\langle\hat{b}^\dag_j\hat{b}_j\rangle_c=\langle\hat{b}^\dag_j\hat{b}_j\rangle_0
exp(-\gamma t)$. At times $t\gg1/\gamma$, we have the vanished
average photon number in mode $b$. This implies that the steady
state of the cavity is the vacuum state in the $b$ basis. In terms
of the original field modes, this procedure means that the atoms
pump in phase only the two-mode squeezed state. So we have the
following field state at steady state
\begin{equation}
\hat{\rho}_c^{ss}=|0,0 \rangle_b\langle
0,0|=\hat{S}^\dag_{12}(\epsilon)|0,0 \rangle_a\langle
0,0|\hat{S}_{12}(\epsilon).
\end{equation}
This is a two-mode squeezed state, whose degree of squeezing
$\epsilon$ is mainly determined by the ratios
$\frac{|\tilde{\Omega}_1|}{|\tilde{\Omega}_2|}$ and
$\frac{|\Delta_1|}{|\Delta_2|}$.

We denote the field quadratures for the cavity modes as
$\hat{X}_i=\frac{1}{2}(\hat{a}_i+\hat{a}^\dag_i)$ and
$\hat{P}_i=-\frac{i}{2}(\hat{a}_i-\hat{a}^\dag_i)$, respectively.
Then the variances in the sum and difference operators in the state
(7) are $V(X_1\pm X_2)=V(P_1\mp P_2)=\frac{1}{2}exp[\pm
2\tanh^{-1}(r)]$. Therefore, two-mode squeezing, i.e.,
Einstein-Podolsky-Rosen (EPR) correlations are established between
the cavity modes at steady state\cite{epr,pra-64-022321}. This state
is reached independently of the the initial state of the cavity
modes, given that each step is implemented for a sufficiently long
time $T$.

It is necessary to analyze the proposal requirements. To realize
this scheme, it needs changes in the intensities of the pump fields
and the transition frequencies of the atoms. It is fairly easy to
tune the intensities of the pump fields in experiments. To change
the transition frequencies of the atoms, an external static field
can be utilized. This protocol needs neither the preparation of the
initial state of the atoms nor the initial state of the cavity. The
whole system has only to stay in the ground states initially. It
does not need the atomic detection nor control of the atomic
velocities and numbers either. The atomic spontaneous emission is
strongly suppressed during the interaction with the cavity modes due
to large atom-field detunings. On the other hand, dissipation of the
cavity field should be negligible in the experiments to implement
the proposal.

We consider some experimental matters. For a potential experimental
system and set of parameters in microwave
resonators\cite{RMP-73-565}, the atomic configuration could be
realized in Rydberg atoms. Typically, alkali atoms $^{85}\mbox{Rb}$
are good candidate\cite{APL}. Then the ground states $\vert
g\rangle$ and $\vert h \rangle$ correspond to the states $\vert
52D_{5/2}\rangle$ hyperfine levels\cite{APL}. The degeneracy of the
states can be lift by using an external magnetic field. These states
are coupled by via Raman transitions involving circularly polarized
cavity modes and pump fields in a resonator.  We choose the single
photon dipole coupling strength as, $g_1\sim g_2=g/(2\pi)\sim 50$
kHZ, and assume laser Rabi frequencies $\Omega_1/(2\pi)\sim 40$ kHZ,
$\Omega_2\sim\Omega_1/0.48$, and atomic excited detunings
$|\Delta_1|/(2\pi)\sim 1$ MHZ, and $|\Delta_2|=2|\Delta_1|$. With
these parameters we obtain the Raman transition rates
$\Theta_1/(2\pi)\sim 2$ kHZ, $\Theta_2/(2\pi)\sim 2.3$ kHZ, and the
squeezing parameter $\epsilon\sim 1.8 (r\sim 0.95)$. The occupation
of the excited state $\vert e\rangle$ can be estimated to be
$\langle \vert e\rangle \langle e\vert\rangle\sim
|\Omega_1/\Delta_1|^2$. Spontaneous emission from the excited state
at a rate $\gamma_e$ thus leads to effective decay rate
$\Gamma_e=|\Omega_1/\Delta_1|^2\gamma_e$. With the given parameters
one can estimate a negligible effect of the spontaneous decay of the
atoms on the fidelity of the squeezed states.

\begin{figure}[h]
\centering
\includegraphics[bb=35 130 588 527,totalheight=2in,clip]{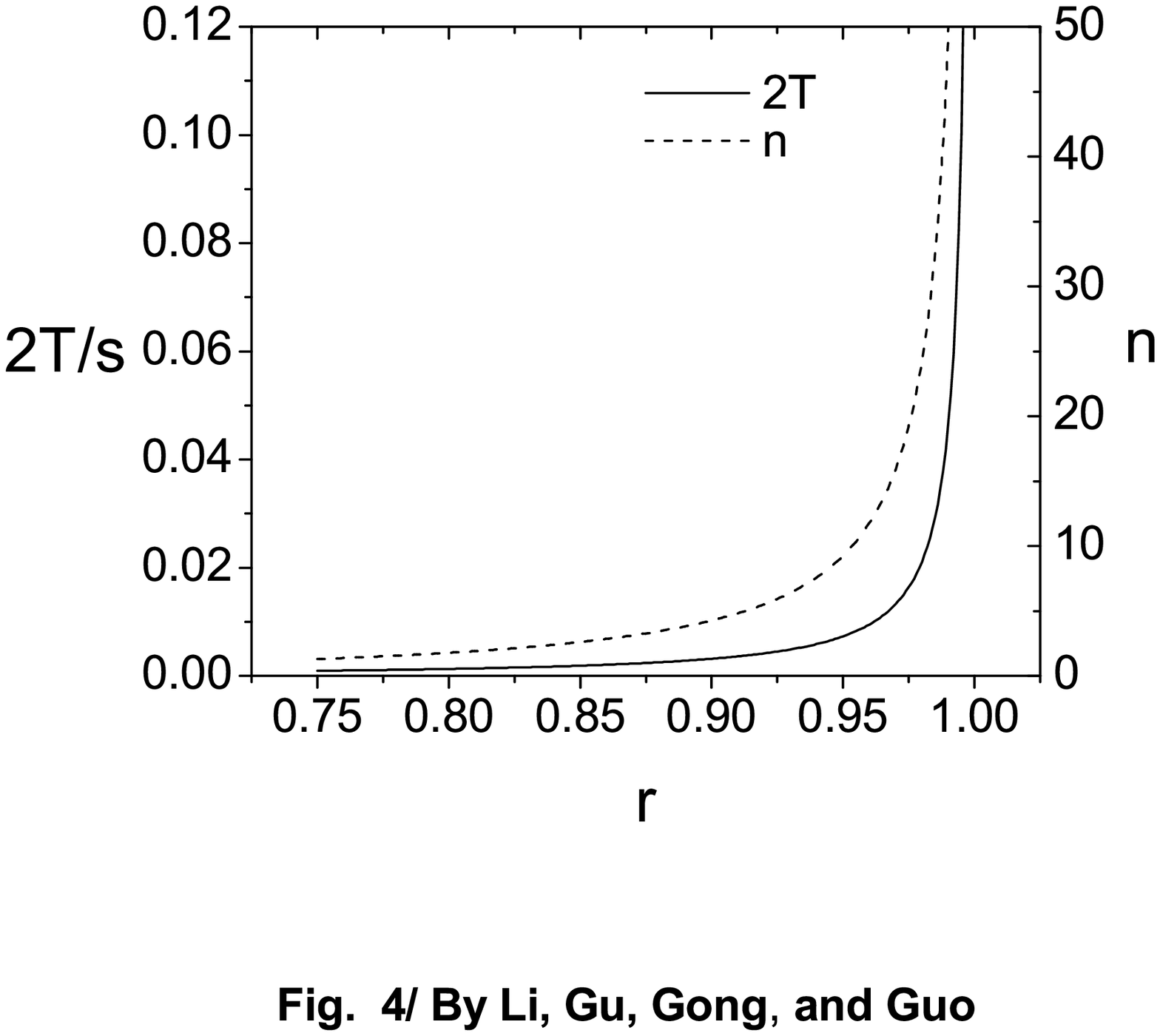}
\caption{ Total preparation time 2T and average photon number n vs
the parameter $r$. }
\end{figure}
The time $T$ for each step to reach
$\langle\hat{b}^\dag_j\hat{b}_j\rangle_T\sim \bar{n}_\infty$ depends
on the initial value $\langle\hat{b}^\dag_j\hat{b}_j\rangle_0=:
\bar{n}_0$, i.e., $T=\gamma^{-1}|\ln (\bar{n}_\infty/\bar{n}_0)|$.
Here $\bar{n}_0=r^2/(1-r^2)$ when the cavity modes are in the vacuum
state initially. For the degree of squeezing $\epsilon\sim
1.83(r\sim0.95)$, from the above parameters we have a total
experimental time $2T\sim 7$ ms, in case of an initially empty
resonator. This result is in line with the currently experimental
setups. Resonators stable over $100$ ms have been reported
recently\cite{APL}. In Fig. 2 we plot the estimated total time to
prepare the squeezed states and the corresponding average photon
number per mode at steady state as a function of the parameter $r$.
From the figure it can be seen that for the squeezing degree
$\epsilon\sim 1.83(r\sim0.95)$, at steady state one can obtain an
average number of 9 photons per cavity mode. This needs about 7 ms
to produce the two-mode squeezing, with a fidelity $F\sim 0.99$.

It is noted that the present proposal is valid on conditions that
the cavity modes should not decay during the experiment. We discuss
this scheme in the weak coupling regime, but one also can get the
same results in the strong coupling conditions by randomizing the
interaction times. This protocol has two distinct advantages. It
does not need the preparation of the initial states of the atoms and
the resonators. All the atoms have only to stay in the ground states
before entering the empty cavity. This offers an convenience in the
experiments. Another advantage is that the atomic decay can hardly
influence this scheme. The two-channel Raman excitations offer the
other convenience in the experiments.


In conclusion, we have proposed a scheme for the generation of a
two-mode field squeezed state in high-Q resonators. This proposal
relies on a form of quantum reservoir engineering and the
two-channel Raman excitations of a beam of three-level atoms. It
does not need neither the preparation of the initial state of the
atoms nor the initial state of the cavity. It is shown that by
suitably choosing the intensities and detunings of fields the
dynamical processes can be selective, which can be utilized to
generate two-mode squeezing between the cavity modes at steady
state. This protocol is robust against atomic spontaneous decay and
can be realized with presently available experimental setups in
cavity QED.

This work was supported by the National Natural Science Foundation
of China under Grants Nos. 10674009, 10334010, 10521002, 10434020
and National Key Basic Research Program No.2006CB921601. Pengbo Li
acknowledges the quite useful discussions with Hongyan Li.


\newpage

\end{document}